\documentclass[oldversion]{aa}
\usepackage{graphicx}
\usepackage{txfonts}
\usepackage{natbib}
\usepackage{amssymb}
\newfont{\gwpfont}{cmssq8 scaled 1000}
\newcommand{\rexcess}{{\gwpfont REXCESS}}

\newcommand{\thetas}{\theta_{\rm s}}
\newcommand{\yo}{y_{\rm o}}	
\newcommand{\jnu}{j_\nu}		
\newcommand{\Psit}{\Psi_{\thetas}}
\newcommand{\Ts}{T_{\thetas}}
\newcommand{\sigt}{\sigma_{\thetas}}
\newcommand{\Ft}{F_{\thetas}}
\newcommand{\nui}{\nu_i}
\def\etal{et al.}
\def\Mv {M_{\rm 500}}
\def\Rv {r_{\rm 500}}
\def\LX {L_{\rm 500}}
\def\Yv {Y_{\rm 500}}
\def\LM  {$\LX$--$\Mv$}
\def\YM  {$\Yv$--$\Mv$}
\def\YL  {$\Yv$--$\LX$}
\def\gtrsim{\mathrel{\hbox{\rlap{\hbox{\lower4pt\hbox{$\sim$}}}\hbox{$>$}}}}

\begin{document}
   \title{The galaxy cluster $Y_{SZ}$-$L_X$ and $Y_{SZ}$-$M$ relations \\ from the WMAP 5-yr data}


   \author{J.-B. Melin\inst{1}
          \and
          J. G. Bartlett\inst{2}
          \and
          J. Delabrouille\inst{3}
          \and
          M. Arnaud\inst{4}
          \and
          R. Piffaretti\inst{4}
          \and
          G. W. Pratt\inst{4}
          }

   \institute{DSM/Irfu/SPP, CEA/Saclay, F-91191 Gif-sur-Yvette Cedex\\
              \email{jean-baptiste.melin@cea.fr}
         \and
             APC -- Universit\'e Paris Diderot, 10 rue Alice Domon et L\'eonie Duquet, 75205 Paris Cedex 13, France\\
            \email{bartlett@apc.univ-paris7.fr}
          \and
            APC -- CNRS, 10 rue Alice Domon et L\'eonie Duquet, 75205 Paris Cedex 13, France\\
            \email{delabrouille@apc.univ-paris7.fr}
          \and
            DSM/Irfu/SAp, CEA/Saclay, F-91191 Gif-sur-Yvette Cedex\\
              \email{monique.arnaud@cea.fr, rocco.piffaretti@cea.fr, gabriel.pratt@cea.fr}
             }

  \date{Received ; accepted }
     
 
  \abstract
{We use multifrequency matched filters to estimate, in the WMAP 5-year data, the Sunyaev--Zel'dovich (SZ) fluxes of 893 ROSAT NORAS/REFLEX clusters spanning the luminosity range $L_{X,[0.1-2.4]\ {\rm keV}} = 2 \times 10^{41} - 3.5\times 10^{45}$ erg s$^{-1}$.  The filters are spatially optimised by using the universal pressure profile recently obtained from combining XMM-Newton  observations of the \rexcess\ sample and numerical simulations. Although the clusters are individually only marginally detected, we are able to firmly measure the SZ signal ($>10\,\sigma$) when averaging the data in luminosity/mass bins. The comparison between the bin-averaged SZ signal versus luminosity and X-ray model predictions shows excellent agreement, implying that there is no deficit in SZ signal strength relative to expectations from the X-ray properties of  clusters.  Using the individual cluster SZ flux measurements, we directly constrain the $Y_{500}-L_{\rm X}$ and $Y_{500}-M_{500}$ relations, where  $Y_{500}$ is the Compton y--parameter integrated over a sphere of radius $r_{500}$. The $Y_{500}$--$M_{500}$ relation, derived for the first time in such a wide mass range, has a normalisation $Y^*_{500}=\left [ 1.60 \pm 0.19 \right ]\times 10^{-3} \, {\rm arcmin}^2$ at $M_{500}=3\times 10^{14}\,h^{-1}\,M_\odot$,  in excellent agreement with the X-ray prediction of $1.54 \times 10^{-3} \, {\rm arcmin}^2$, and a mass exponent of  $\alpha=1.79 \pm 0.17$, consistent with the self-similar expectation of $5/3$. Constraints on the redshift exponent are weak due to the limited redshift range of the sample, although they are compatible with self-similar evolution.}
  

   \maketitle
   
%

\section{Introduction}

Capability to observe the Sunyaev-Zel'dovich (SZ) effect has improved immensely in recent years. Dedicated instruments now produce high resolution images of single objects \citep[e.g.][]{kit04,halversonetal09,nor09} and moderately large samples of high--quality SZ measurements of previously--known clusters \citep[e.g.,][]{mro09,pla09}. In addition, large-scale surveys for clusters using the SZ effect are underway, both from space with the Planck mission \citep[]{2007NewAR..51..287V,2003NewAR..47.1017L} and from the ground with several dedicated telescopes,  such as the South Pole Telescope \citep{car09} leading to the first discoveries of clusters solely through their SZ signal \citep{stanisetal09}. These results open the way for a better understanding of the $SZ-Mass$ relation and, ultimately, for cosmological studies with large SZ cluster catalogues.

The SZ effect probes the hot gas in the intracluster medium (ICM).  Inverse Compton scattering of cosmic microwave background (CMB) photons by free electrons in the ICM creates a unique spectral distortion \citep{sun70,sun72} seen as a frequency--dependent change in the CMB surface brightness in the direction of galaxy clusters that can be written as $\Delta i_\nu(\hat{n}) = y(\hat{n}) j_\nu(x)$, where $j_\nu$ is a universal function of the dimensionless frequency $x=h\nu/kT_{\rm cmb}$. The Compton $y$--parameter is given by the integral of the electron pressure along the line--of--sight in the direction $\hat{n}$,
\begin{equation}
y = \int_{\hat{n}}\; \frac{kT_e}{m_e c^2} n_e\sigma_T dl, 
\end{equation}
where $\sigma_T$ is the Thomson cross section.

Most notably, the integrated SZ flux from a cluster directly measures the total thermal energy of the gas.  Expressing this flux in terms of the integrated Compton $y$--parameter $Y_{\rm SZ}$ -- defined by $\int\, d\Omega\,\Delta\,i_\nu(\hat{n})\,=\,Y_{\rm SZ} j_\nu(x)$ -- we see that $Y_{\rm SZ}\,\propto\,\int\,d\Omega\,dl n_eT_e\propto \int\,n_eT_e dV$.  For this reason, we expect $Y_{\rm SZ}$ to closely correlate with total cluster mass, $M$, and to provide a low--scatter mass proxy.  

This expectation, borne out by both numerical simulations \citep[e.g.,][]{das04,mot05,kra06} and indirectly from X--ray observations using $Y_{\rm X}$, the product of the gas mass and mean temperature \citep[][]{2007ApJ...668....1N,2007A&A...474L..37A,vik09}, strongly motivates the use of SZ cluster surveys as cosmological probes. 
Theory predicts the cluster abundance and its evolution -- the mass function -- in terms of $M$ and the cosmological parameters.  With a good mass proxy, we can measure the mass function and its evolution and hence constrain the cosmological model, including the properties of dark energy.  In this context the relationship between the integrated SZ flux and total mass, $Y_{\rm SZ}-M$, is fundamental as the required link between theory and observation.  Unfortunately, despite its importance, we are only beginning to observationally constrain the relation  \citep{bonamente08,marrone09}. 

Several authors have extracted the cluster SZ signal from WMAP data \citep{2003ApJS..148....1B, 2007ApJS..170..288H, 2009ApJS..180..225H}. However, the latter are not ideal for SZ observations:  the instrument having been designed  to measure primary CMB anisotropies on scales larger than galaxy clusters, the spatial resolution and sensitivity of the sky maps render cluster detection difficult.  Nevertheless, these authors extracted the cluster SZ signal by either cross--correlating with the general galaxy distribution \citep{2003ApJ...597L..89F, 2004MNRAS.347L..67M, 2004ApJ...613L..89H, 2006A&A...449...41H} or `stacking' existing cluster catalogues in the optical or X-ray \citep{2006ApJ...648..176L, 2007MNRAS.378..293A, 2008ApJ...675L..57A, bie07, 2009MNRAS.tmp.1927D}.  These analyses indicate that
an isothermal $\beta$-model is not a good description of the SZ profile, and some suggest that the SZ signal strength is lower than expected from the X-ray properties of the clusters \citep{2006ApJ...648..176L, bie07}.

Recent in--depth  X-ray studies of the ICM pressure profile demonstrate regularity in shape and simple scaling with cluster mass.  Combining these observations with numerical simulations leads to a universal pressure profile \citep{2007ApJ...668....1N,arnaud09} that is best
fit by a modified NFW profile.  The isothermal $\beta$--model, on the other hand, does not provide an adequate fit. 
From this newly determined X--ray pressure profile, we can infer the expected SZ profile, $y(r)$, and the $Y_{\rm SZ}-M$ relation at low 
redshift  \citep{arnaud09}.

It is in light of this recent progress from X--ray observations that we present a new analysis of the SZ effect in WMAP with the aim of 
constraining the SZ scaling laws.  
We build a multifrequency matched filter \citep{2002MNRAS.336.1057H, 2006A&A...459..341M} based on the known spectral shape of the thermal SZ effect and the shape of the universal pressure profile of \citet{arnaud09}. This profile was derived from \rexcess\ \citep{boe07}, a sample expressly designed to measure the structural and scaling properties of the local X--ray cluster population by means of an unbiased, representative sampling in luminosity. Using the multifrequency matched filter, we search for the SZ effect in WMAP from a catalogue of 893 clusters detected by ROSAT, maximising the signal--to--noise by adapting the filter scale  to the expected characteristic size of each cluster. The size is estimated through the luminosity--mass relation derived from the \rexcess\ sample by \citet{2009A&A...498..361P}. 

We then use our SZ measurements to directly determine the $Y_{\rm SZ}-L_X$ and $Y_{\rm SZ}-M$ relations and compare to expectations based on the universal X--ray pressure profile. As compared to the previous analyses of \citet{bonamente08} and \citet{marrone09}, the large number of systems in our WMAP/ROSAT sample allows us to constrain both the normalisation and slope of the $Y_{\rm SZ}-L_X$ and $Y_{\rm SZ}-M$ relations over a wider mass range and in the larger  aperture of $r_{500}$.  In addition, the analysis is based on a more realistic pressure profile than in these analyses, which were based on an isothermal $\beta$--model.  Besides providing a direct constraint on these relations, the good agreement with X--ray predictions implies that there is in fact no deficit in SZ signal strength relative to expectations from the X-ray properties of these clusters.

The discussion  proceeds as follows.  We first present the WMAP 5--year data and the ROSAT cluster sample used, a combination
of the REFLEX and NORAS catalogues. We then present the SZ model 
based on the X--ray--measured pressure profile (Sec.~\ref{cluster_model}). In Sec.~\ref{sz_flux}, we discuss our SZ measurements, after first describing how we extract the signal 
using the matched filter. Section~\ref{err_bud}  details the error budget.  We  compare our measured scaling laws to the X--ray predictions in Sect.~\ref{yl_relation} and~\ref{ym_relation} and then conclude  in Sec.~\ref{discussion}.  Finally, we collect useful SZ definitions
and unit conversions in the Appendices.

Throughout this paper, we use the WMAP5--only cosmological parameters set as our `fiducial cosmology', i.e. $h=0.719$, $\Omega_M=0.26$, $\Omega_\Lambda=0.74$, where  $h$ is the Hubble parameter at redshift $z\,=\,0$ in units of $100\,$\,km/s/Mpc. We note $h_{70} = h/0.7$ and $E(z)$ is the Hubble parameter at redshift $z$ normalised to  its present value.  $\Mv$ is defined as the mass within  the radius $\Rv$ at which the mean mass density is 500 times the critical density, $ \rho_{crit}(z)$, of the universe at the cluster redshift: $M_{500} = {4 \over 3} \pi  \, \rho_{crit}(z) \, 500 \, r_{500}^3$.

\section{The WMAP-5yr data and the NORAS/REFLEX cluster sample}
\label{data}

\subsection{The WMAP-5yr data}

We work with the WMAP full resolution coadded five year sky temperature maps at each frequency channel (downloaded from the LAMBDA archive\footnote{http://lambda.gsfc.nasa.gov/}). There are five full sky maps corresponding to frequencies 23, 33, 41, 61, 94 GHz (bands K, Ka, Q, V, W respectively). The corresponding beam full widths at half maximum are approximately 52.8, 39.6, 30.6 21.0 and 13.2 arcmins.
The maps are originally at HEALPix\footnote{http://healpix.jpl.nasa.gov} resolution nside=512 (pixel= 6.87 arcmin). Although this is reasonably adequate to sample WMAP data, it is not adapted to the multifrequency matched filter algorithm we use to extract the cluster fluxes. We oversample the original data, to obtain nside=2048 maps,  by zero-padding in harmonic space. In detail, this is performed by computing the harmonic transform of the original maps, and then performing the back transform towards a map with nside=2048, with a maximum value of $\ell$ of $\ell_{\rm max}$ =  750, 850, 1100, 1500, 2000 for the K, Ka, Q, V, W bands respectively. The upgraded maps are smooth and do not show pixel edges as we would have obtained using the HEALPix upgrading software, based on the tree structure of the HEALPix pixelisation scheme. This smooth upgrading scheme is important as the high spatial frequency content induced by pixel edges would have been amplified through the multifrequency matched filters implemented in harmonic space. 

In practice, the multifrequency matched filters are implemented locally on small, flat patches (gnomonic projection on tangential maps), which permits adaptation of the filter to the local conditions of noise and foreground contamination. We divide the sphere into 504 square tangential overlapping patches (100 ${\rm deg}^2$ each, pixel=1.72 arcmin). All of the following analysis is done on these sky patches.

The implementation of the matched filter requires knowledge of the WMAP beams. In this work, we assume symmetric beams, for which the transfer function $b_\ell$ is computed, in each frequency channel, from the noise-weighted average of the transfer functions of individual differential assemblies (a similar approach was used in \citealt{2009A&A...493..835D}).

\begin{figure}[t]
   \centering
   \includegraphics[scale=0.5]{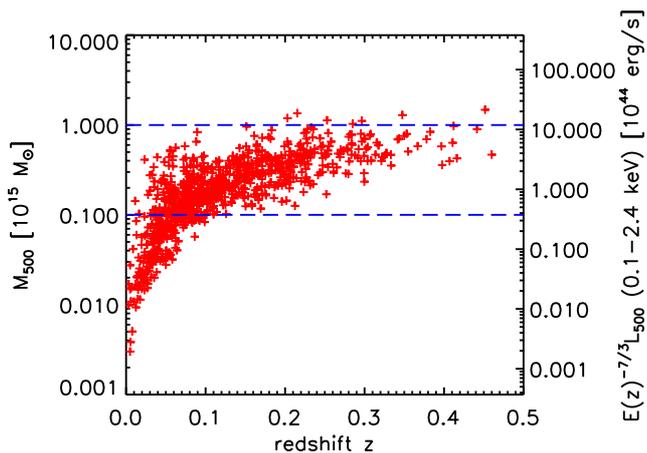}
   \caption{Inferred masses for the 893 NORAS/REFLEX clusters as a function of redshift. The cluster sample is flux limited. The right vertical axis gives the corresponding X-ray luminosities scaled by $E(z)^{-7/3}$. The dashed blue lines delineate the mass range over which the $L_{500}$-$M_{500}$ relation from~\citet{2009A&A...498..361P} was derived.}
    \label{fig:m_z_relation}
    \end{figure}

\subsection{The NORAS/REFLEX cluster sample and derived X--ray properties}
\label{rass}

We construct our cluster sample from the largest published X-ray catalogues: NORAS \citep{2000ApJS..129..435B} and REFLEX \citep{2004A&A...425..367B}, both constructed from the ROSAT All-Sky Survey. We merge the cluster lists given in Tables 1, 6 and 8 from ~\citet{2000ApJS..129..435B}  and Table 6 from~\citet{2004A&A...425..367B} and since the luminosities of the NORAS clusters are given in a standard cold dark matter (SCDM) cosmology ($h=0.5$, $\Omega_M=1$), we converted them to the WMAP5 cosmology.  We also convert the luminosities of REFLEX  clusters from the basic $\Lambda$CDM cosmology ($h=0.7$, $\Omega_M=0.3$, $\Omega_\Lambda=0.7$) to the more precise WMAP5 cosmology. Removing clusters appearing in both catalogues leaves 921 objects, of which 893 have measured redshifts. We use these 893 clusters in the analysis detailed in the next Section.

The NORAS/REFLEX luminosities $L_{\rm X}$, measured in the soft $[0.1$--$2.4]\,{\rm keV}$ energy band,  are given within various apertures depending on the cluster. We convert the luminosities $L_{\rm X}$ to $L_{500}$, the luminosities within $\Rv$, using an iterative scheme. This scheme is based on the mean electron density profile of the \rexcess\ cluster sample \citep{cro08},  which allows conversion of the luminosity between various apertures,  and  the \rexcess\ $L_{500}$--$M_{500}$ relation \citep{2009A&A...498..361P},  which implicitly relates $\Rv$ and $L_{500}$. The procedure thus simultaneously yields an estimate of the cluster mass, $M_{500}$, and the corresponding angular extent $ \theta_{500} = \Rv/D_{\rm ang}(z)$, where  $D_{ang}(z)$ is the angular distance at redshift $z$.   In the following we consider values derived from relations both corrected and uncorrected for  Malmquist bias. The relations are described by  the following power law models\footnote{Since we consider a standard self-similar model, we used the power law relations given in Appendix B of \citet{arnaud09}. They are derived as in \citet{2009A&A...498..361P}  with the same luminosity data but  for  masses derived from a standard slope $M_{500}$--$Y_{\rm X}$ relation.}:
\begin{equation}
\label{lx_m500}
     E(z)^{-7/3} \, L_{500} = C_M \, \left ( M_{500} \over 3 \times 10^{14} h_{70}^{-1} M_\odot \right )^{\alpha_M}
\end{equation}
where the normalisation $C_M$, the exponent $\alpha_M$ and the dispersion (nearly constant with mass) are given in Table~\ref{tab:lx_m_param}.
The $L_{500}$--$M_{500}$ relation was  derived in the mass range $[10^{14}$--$10^{15}]\,M_\odot$. These limits are shown in Fig.~\ref{fig:m_z_relation}. Note that we assume the relation is valid for lower masses.

     \begin{table}
      \caption[]{Values for the parameters of the $L_X-M$ relation derived from \rexcess\  data \citep{2009A&A...498..361P,arnaud09}}
         \label{tab:lx_m_param}
        \begin{center}
        \begin{tabular}{cccc}
         \hline
         \hline
          Corrected for MB & $\log \left ( C_M \over 10^{44} h_{70}^{-2} {\rm [erg s^{-1}]} \right ) $ & $\alpha_M$ & $\sigma_{\log L - \log M}$  \\
            \hline
              no &  0.295 & 1.50 & 0.183 \\
               yes & 0.215 & 1.61 & 0.199 \\
          \hline
        \end{tabular}
        \end{center}
   \end{table}

 The final catalogue of 893 objects contains the position of the clusters (longitude and latitude), the measured redshift $z$, the derived X-ray luminosity $L_{500}$, the mass $M_{500}$ and the angular extent $ \theta_{500}$. The clusters uniformly cover the celestial sphere at Galactic latitudes above $|b|>20 \, {\rm deg}$. Their luminosities $L_{500}$ range from 0.002 to $35.0 \, 10^{44} {\rm erg/s}$, and their redshifts from 0.003 to 0.460. Figure~\ref{fig:m_z_relation} shows the masses $M_{500}$ as a function of redshift $z$ for the cluster sample (red crosses). The corresponding corrected luminosities $L_{500}$ can be read on the right axis. The typical luminosity correction from measured $L_X$ to $L_{500}$ is about $10\%$.  The progressive mass cut-off with redshift (only the most massive clusters are present at high $z$) reflects the flux limited nature of the sample.

\section{The cluster SZ model}
\label{cluster_model}

In this Section we describe the cluster SZ model,  based on X-ray observations of the \rexcess\ sample combined with numerical simulations, as presented  in \citet{arnaud09}. We use the standard self-similar model presented in their Appendix B.  Given a cluster mass $M_{500}$ and redshift $z$, the model predicts the electronic pressure profile. This gives both the SZ profile shape and  $Y_{500}$, 
 the SZ flux integrated in a sphere of radius $r_{500}$.
 \subsection{Cluster shape}

The  dimensionless universal pressure profile is taken from Eq.~ B1 and Eq.~B2  of \citet{arnaud09}:

\begin{equation}
\label{cluster_profile}
     {P(r) \over P_{500}} = {P_0 \over x^\gamma (1+x^\alpha)^{(\beta-\gamma)/\alpha}}
\end{equation}
where $x=r/r_s$ with $r_s=r_{500}/c_{500}$ and $c_{500}=1.156$, $\alpha=1.0620$, $\beta=5.4807$, $\gamma=0.3292$ and with $P_{500}$  defined in Eq.~\ref{p500} below.

This profile shape is used to optimise the SZ signal detection. As described below in Sect.~\ref{sz_flux}, we extract the $Y_{SZ}$ flux from WMAP data  for each ROSAT system fixing   $c_{500}$,  $\alpha$, $\beta$, $\gamma$ to the above values, but leaving the normalisation free.

\subsection{Normalisation}

\begin{figure*}[t]
   \centering
   \includegraphics[scale=0.5]{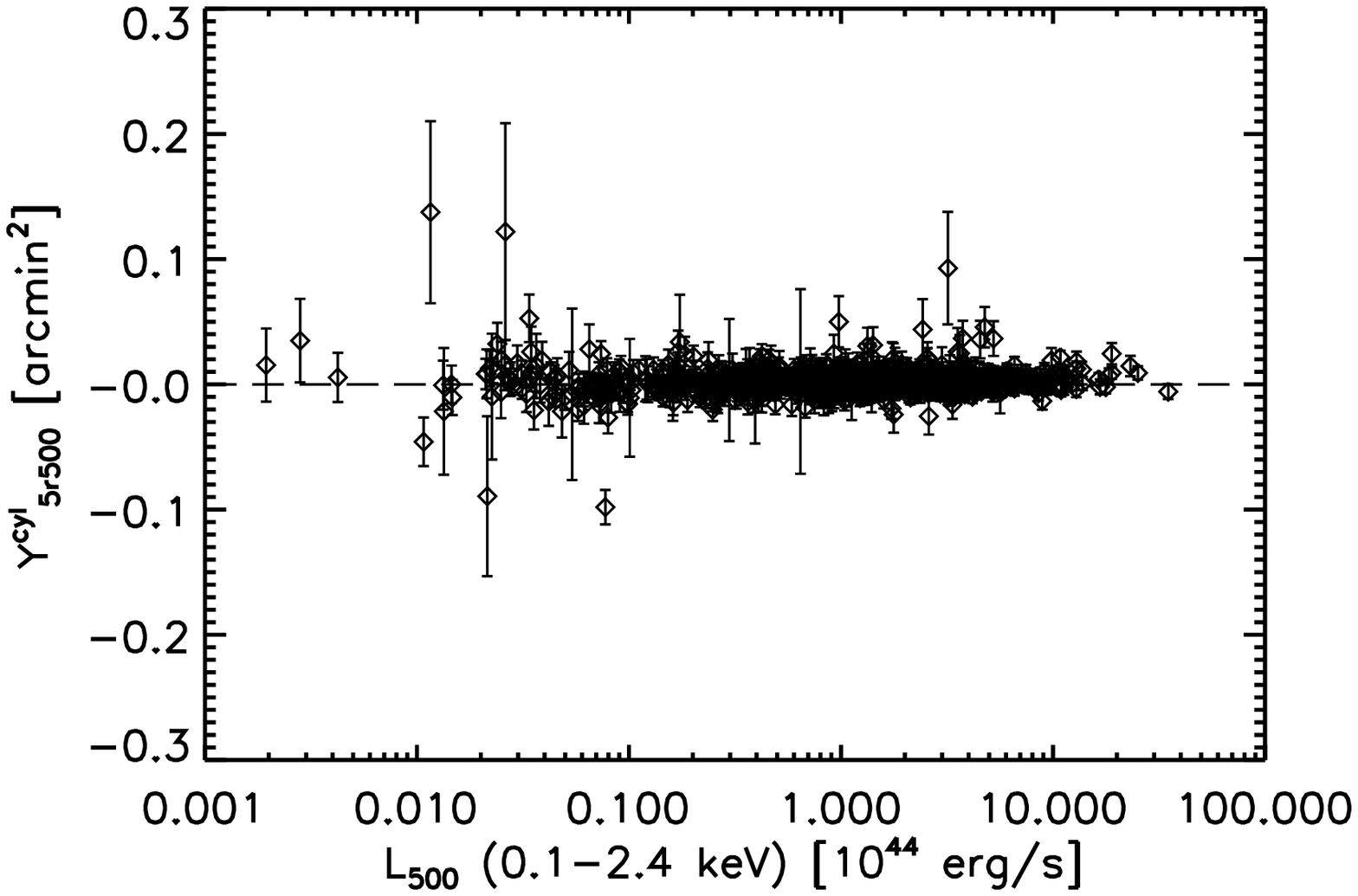}
   \includegraphics[scale=0.5]{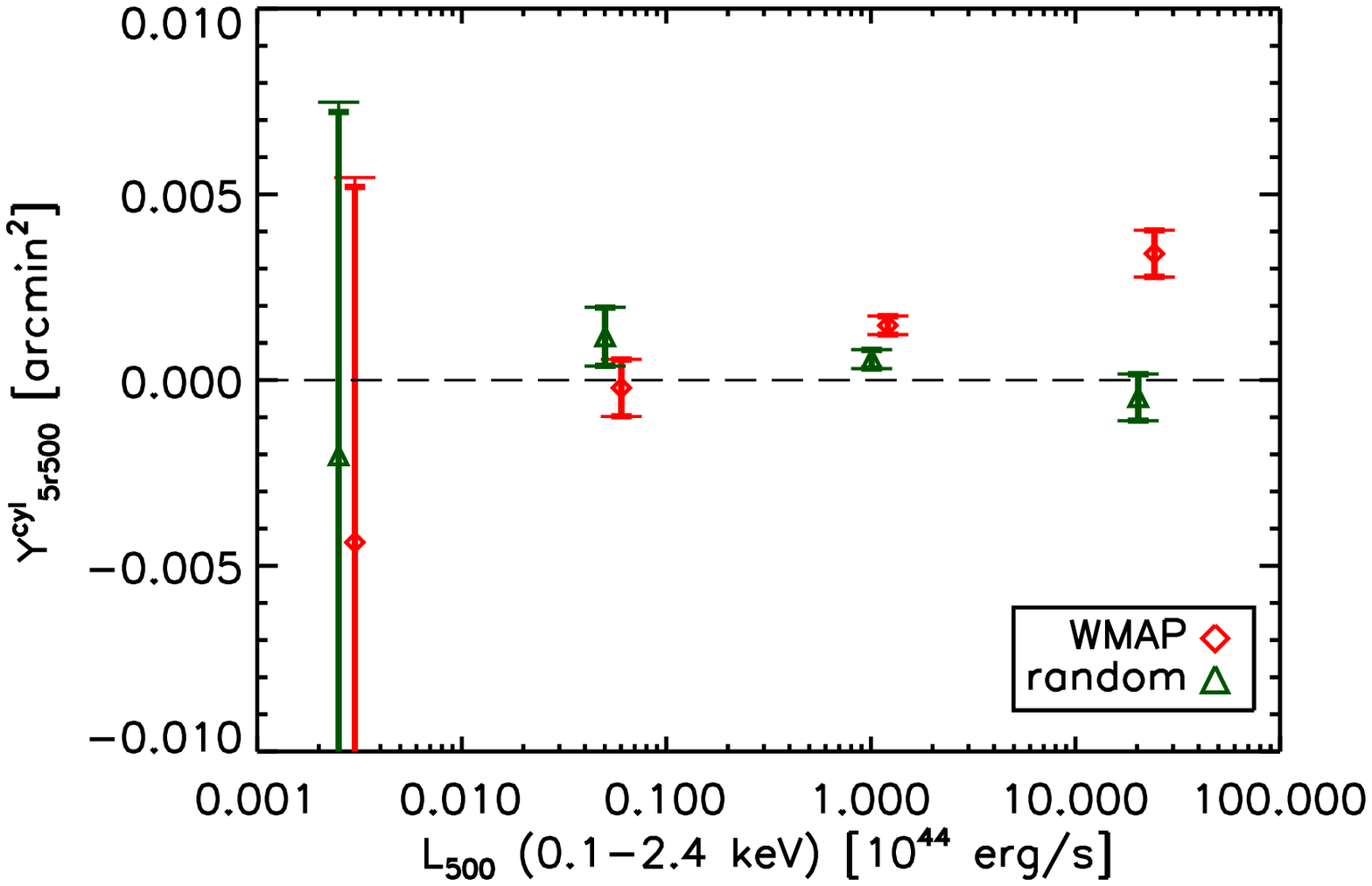}
   \caption{{\it Left:} Estimated SZ flux from a cylinder of aperture radius $5 \times \Rv$ ($Y^{cyl}_{5r500}$) as a function of the X-ray luminosity in an aperture of $r_{500}$ ($L_{500}$), for the 893 NORAS/REFLEX clusters. The individual clusters are barely detected. The bars give the total 1 $\sigma$ error. {\it Right:} Red diamonds are the weighted average signal in 4 logarithmically--spaced luminosity bins.  The two high luminosity bins exhibit significant SZ cluster flux. Note that we have divided the vertical scale by 30 between Fig. left and right. The thick and thin bars give the 1 $\sigma$ statistical and total errors, respectively. Green triangles (shifted up by 20\% with respect to diamonds for clarity) show the result of the same analysis when the fluxes of the clusters are estimated at random positions instead of true cluster positions. }
              \label{fig:cy_lx_raw_binned}
    \end{figure*}

The model allows us to compute the physical pressure profile as a function of mass and $z$,  thus the $Y_{SZ}$-$M_{500}$ relation by integration of  $P(r)$ to $r_{500}$. For the shape parameters given above,  the normalisation  parameter $P_0=8.130 \, h_{70}^{3/2} = 7.810$ and the self-similar definition of 
$P_{500}$ \citep[][Eq.~5 and Eq.~B2]{arnaud09},
\begin{equation}
\label{p500}
     P_{500} = 1.65 \times 10^{-3} E(z)^{8/3} \left ( {M_{500} \over 3\times10^{14}\, h_{70}^{-1} M_\odot} \right )^{2/3}  h_{70}^2  \; {\rm keV \, cm^{-3}},
      \end{equation}
one obtains:
     \begin{equation}
     \label{ym_ss_rel}
      Y_{500} \, [{\rm arcmin}^2]= Y^*_{500} \, \left ({M_{500} \over 3\times10^{14}\,h^{-1}\,M_\odot} \right )^{5/3} \, E(z)^{2/3} \, \left ({D_{ang}(z) \over 500 \, {\rm Mpc}} \right )^{-2},
 \end{equation}
\noindent where $Y^*_{500}=1.54\times10^{-3} \, \left ({h \over 0.719} \right )^{-5/2} \, {\rm arcmin}^2$.  Equivalently, one can write:
\begin{equation}
      Y_{500} \, \, [{\rm Mpc}^2] = Y^*_{500} \; \left ({M_{500} \over 3\times10^{14}\,h^{-1}\,M_\odot} \right )^{5/3} \; E(z)^{2/3}     
      \end{equation}
\noindent where $Y^*_{500}=3.27\times10^{-5} \, \left ({h \over 0.719} \right )^{-5/2} \, {\rm Mpc}^2$. Details of unit conversions are given in Appendix~\ref{sz_conv}.
The mass dependence ($M_{500}^{5/3}$) and the redshift dependence ($E(z)^{2/3}$) of the relation are self-similar by construction. This model is used to predict the $Y_{500}$  value for each cluster.  These predictions are compared to the WMAP-measured values in Figs.~\ref{fig:cy_lx_relation}, \ref{fig:cy_m_relation}, \ref{fig:evolution} and \ref{fig:cy_m_loglog}.

\section{Extraction of the SZ flux}
\label{sz_flux}

\subsection{Multifrequency Matched Filters}
\label{mmf_sec}

We use multifrequency matched filters to estimate cluster fluxes from the WMAP frequency maps. By incorporating prior
knowledge of the cluster signal, i.e.,  its spatial and spectral 
characteristics, the method maximally enhances the signal--to--noise of a SZ cluster source by optimally filtering the data. The universal profile shape described in Sec.~\ref{cluster_model} is assumed, and we evaluate the effects of uncertainty in this profile as outlined in Sec.~\ref{err_bud} where we discuss our overall error budget.  We fix the position and the characteristic radius $\thetas$ of each cluster and estimate only its flux.    The position is taken from the NORAS/REFLEX catalogue and $\thetas=\theta_{500}/c_{500}$  with  $\theta_{500}$ derived from X-ray data as described in Sec.~\ref{rass}. 
Below, we recall the main features of the multifrequency matched filters.  More details can be found in~\citet{2002MNRAS.336.1057H} or~\citet{2006A&A...459..341M}.

Consider a cluster with known radius $\thetas$ and unknown central  $y$--value $\yo$ positioned at a known point $\vec{x}_{\rm o}$  on the sky.  The region is covered by the five WMAP maps $M_i(\vec{x})$ at frequencies $\nui$=23, 33, 41, 61, 94 GHz ($i=1,...,5$).  We arrange the survey maps into a column vector $\vec{M}(\vec{x})$ whose $i^{th}$ component is the map at frequency $\nui$. The maps contain the cluster SZ signal plus noise:
\begin{equation}
\vec{M}(\vec{x}) = \yo\vec{\jnu}\Ts(\vec{x}-\vec{x}_{\rm o}) + \vec{N}(\vec{x})
\end{equation}
where $\vec{N}$ is the noise vector (whose components are noise maps at the different observation frequencies) and $\vec{\jnu}$ is a vector with components given by the SZ spectral function $\jnu$ evaluated at each frequency.  Noise in this context refers to both instrumental noise as well as all signals other than the cluster thermal SZ effect; it thus also comprises astrophysical foregrounds, for example, the primary CMB anisotropy, diffuse Galactic emission and extragalactic point sources.   $\Ts(\vec{x}-\vec{x}_{\rm o})$  is the SZ template, taking into account the WMAP beam,  at projected distance $(\vec{x}-\vec{x}_{\rm o})$ from the cluster centre, normalised to a central value of unity before convolution. It is computed by integrating along the line--of--sight and normalising  the universal pressure profile (Eq.~\ref{cluster_profile}). The profile is truncated at $5 \times r_{500}$ (i.e. beyond the virial radius) so that what is actually  measured is the flux within  a cylinder of aperture radius $5 \times r_{500}$.

X-ray observations are typically well-constrained out to $r_{500}$. Our decision to integrate out to $5 \times r_{500}$ is motivated by the fact that for the majority of clusters the radius $r_{500}$ is of order the Healpix pixel size (nside=512, pixel=6.87 arcmin). Integrating only out to $r_{500}$ would have required taking into account that only a fraction of the flux of some pixels contributes to the true SZ flux in a cylinder of aperture radius $r_{500}$. We thus obtain the total SZ flux of each cluster by integrating out to $5 \times r_{500}$, and then convert this to the value in a sphere of radius $r_{500}$ for direct comparison with the X-ray prediction.

The multifrequency matched filters $\vec{\Psit}(\vec{x})$ return a minimum variance unbiased estimate, $\hat{\yo}$, of
$\yo$ when centered on the cluster:
\begin{equation}
\label{eq:filter}
\hat{\yo} = \int d^2x\; \vec{\Psit}^t(\vec{x}-\vec{x}_{\rm o}) 
      \cdot \vec{M}(\vec{x})
\end{equation}
where superscript $t$ indicates a transpose (with complex conjugation when necessary).  This is just a linear combination of the maps, each
convolved with its frequency--specific filter $(\Psit)_i$.
The result expressed in Fourier space is:
\begin{equation}
\vec{\Psit}(\vec{k}) = \sigt^2 \vec{P}^{-1}(\vec{k})\cdot \vec{\Ft}(\vec{k})
\end{equation}
where 
\begin{eqnarray}
\vec{\Ft}(\vec{k})   & \equiv & \vec{\jnu} \Ts(\vec{k})\\
\label{eq:sigt}
\sigt          & \equiv & \left[\int d^2k\; 
     \vec{\Ft}^t(\vec{k})\cdot \vec{P}^{-1} \cdot
     \vec{\Ft}(\vec{k})\right]^{-1/2}
\end{eqnarray}
with $\vec{P}(\vec{k})$ being the noise power spectrum, a matrix in
frequency space with components $P_{ij}$ defined by $\langle
N_i(\vec{k})N_j^*(\vec{k}')\rangle_N=P_{ij}(\vec{k})
\delta(\vec{k}-\vec{k}')$.    
The quantity $\sigt$ gives the total noise variance through the filter, corresponding to the statistical errors quoted in this paper. The other uncertainties are estimated separately as described in Sec.~\ref{disp_err}. The noise power spectrum $\vec{P}(\vec{k})$ is directly estimated from the maps: since the SZ signal is subdominant at each frequency, we assume $\vec{N}(\vec{x}) \approx \vec{M}(\vec{x})$ to do this calculation. We undertake the Fourier transform of the maps and average their cross-spectra in annuli with width $\Delta l=180$.

\subsection{Measurements of the SZ flux}
 
The derived total WMAP flux from a cylinder of aperture radius $5 \times \Rv$ ($Y^{cyl}_{5r500}$) for the 893 individual NORAS/REFLEX clusters is shown as a function of the measured X--ray luminosity $\LX$ in the left-hand panel of Fig.~\ref{fig:cy_lx_raw_binned}. The clusters are barely detected individually. The average signal--to--noise ratio (S/N) of the total population is 0.26 and only 29 clusters are detected at $S/N>2$, the highest detection being at 4.2. However, one can distinguish the deviation towards positive flux at the very high luminosity end. 

In the right-hand panel of Fig.~\ref{fig:cy_lx_raw_binned}, we average the 893 measurements in four logarithmically--spaced luminosity bins (red diamonds plotted at bin center). The number of clusters are 7, 150, 657, 79 from the lowest to the highest luminosity bin. Here and in the following, the bin average is defined as the weighted mean of the SZ flux  in the bin (weight of $1/\sigt^2$). The thick error bars  correspond to the statistical uncertainties on the WMAP data only,  while the thin bar gives the total errors as discussed in Sec.~\ref{disp_err}.
The SZ signal is clearly detected in the two highest luminosity bins (at 6.0 and 5.4 $\sigma$, respectively). As a demonstrative check, we have undertaken the analysis a second time using random cluster positions. The result is shown by the green triangles in Fig.~\ref{fig:cy_lx_raw_binned}  and is consistent with no SZ signal, as expected.
 
In the following Sections,  we study both the relation between  the SZ signal and the X-ray luminosity  and that with the mass $M_{500}$.  We consider $Y_{500}$,  the SZ flux from a sphere of radius $r_{500}$, converting the measured $Y^{cyl}_{5r500}$ into $Y_{500}$ as described in Appendix~\ref{sz_ref}. This allows a more direct  comparison with the model derived from X--ray observations (Sec.~\ref{cluster_model}).  Before presenting the results, we first discuss the overall error budget.

 \section{Overall error budget}
\label{err_bud}

\subsection{Error due to dispersion in X-ray properties}
\label{disp_err}

The  error $\sigt$ on $\Yv$ given by the multifrequency matched filter only  includes the statistical SZ measurement  error,  due to the instrument (beam, noise) and to the astrophysical contaminants (primary CMB, Galaxy, point sources).  However, we must also  take into account: 1) uncertainties on the cluster mass estimation from the X-ray luminosities via the $L_{500}-M_{500}$ relation, 2) uncertainties on the cluster profile parameters. These are sources of error on individual $\Yv$ estimates (actual parameters for each individual cluster may deviate somewhat from the average cluster model). These deviations from the mean, however, induce additional {\it random} uncertainties on statistical quantities derived from $\Yv$, i.e. bin averaged $Y_{500}$ values and the \YL\ scaling relation parameters.  Their impact on the \YM\ relation, which depends directly on the $\Mv$ estimates, is also an additional random uncertainty.

The uncertainty  on $\Mv$ is dominated  by the intrinsic dispersion in the  \LM\  relation. Its effect is  estimated  by a Monte Carlo (MC) analysis of 100 realisations. We use the  dispersion  at $z=0$ as estimated by  \citet{2009A&A...498..361P}, given in Table~\ref{tab:lx_m_param}. For each realisation, we draw a random mass $\log M_{500}$ for each cluster from a Gaussian distribution with mean given by the  \LM\   relation and standard deviation $\sigma_{\log L - \log M}/\alpha_M$. We then redo the full analysis (up to the fitting of the $Y_{SZ}$ scaling relations) with the new values of $M_{500}$ (thus $\thetas$). 

The second  uncertainty  is due to the observed dispersion in the cluster profile shape, which  depends on radius as shown  in~\citet[][$\sigma_{\log P} \sim 0.10$ beyond the core]{arnaud09}. Using  new 100 MC realisations, we estimate this error by drawing  a cluster profile in the log--log plane from a Gaussian distribution with mean given by Eq.~\ref{cluster_profile} and standard deviation depending on the cluster radius as shown in the lower panel of Fig.~2 in~\citet{arnaud09}. 

The total error on $Y_{500}$ and on the scaling law parameters is calculated from the quadratic sum of the standard deviation of both the above MC  realisations and the error due to the SZ measurement  uncertainty. 
  
  \subsection{The Malmquist bias}
\label{mal_bias}

The NORAS/REFLEX sample is flux limited and is thus subject to the Malmquist  bias (MB). This is a source of systematic error. Ideally we should use  a \LM\  relation which takes into account the specific  bias of the sample, i.e. computed from the true \LM\  relation, with dispersion and  bias according to each survey selection function.  We  have an estimate of the true, ie MB corrected,  \LM\ relation,  from the published analysis of \rexcess\  data (Table 1). However, while the  REFLEX selection function is known and available, this is not the case for the NORAS sample.  This means that we cannot perform a fully consistent analysis. 
In order to estimate the impact of the Malmquist Bias we thus present, in the following, results for two cases. 

In the first case, we use the published  \LM\  relation derived directly from the \rexcess\ data, i.e. not corrected for the \rexcess\ MB (hereafter the \rexcess\  \LM\ relation).  Note that the \rexcess\ is a sub-sample of REFLEX. Using this relation should result in correct masses  if the Malmquist bias for the NORAS/REFLEX sample is the same as that for the \rexcess. The \YM\ relation derived in this case would also be correct and could be consistently compared with  the X--ray predicted relation. We recall that this relation was derived from pressure and mass measurements that are  not sensitive to the Malmquist bias.  However $\LX$ would remain uncorrected so that the  \YL\ relation derived in this case should be viewed as a relation uncorrected for the Malmquist bias.  
In the second case, we use the MB corrected \LM\  relation (hereafter the intrinsic \LM\  relation).  This reduces to assuming that the Malmquist bias is negligible for the NORAS/REFLEX sample. 
The comparison  of the two analyses provides an estimate of the direction and amplitude of the effect of the Malmquist  bias on our results.  The \rexcess\  \LM\ relation is expected to be closer to the  \LM\ relation for the NORAS/REFLEX sample than the intrinsic relation. The discussions and figures correspond to the results obtained when using the former,  unless explicitly specified.
 
 \begin{figure*}[tp]
   \centering
   \includegraphics[scale=0.5]{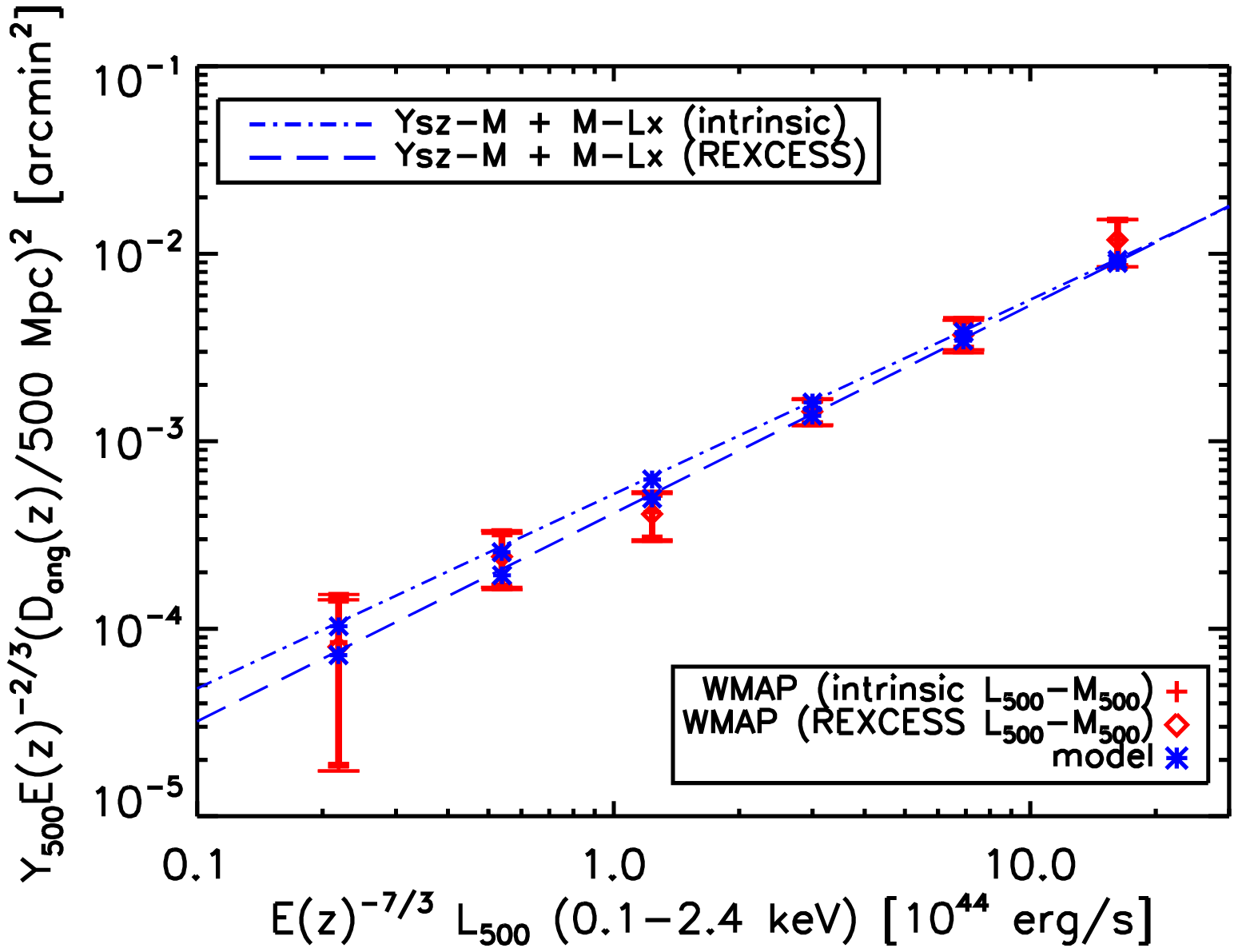}
   \includegraphics[scale=0.5]{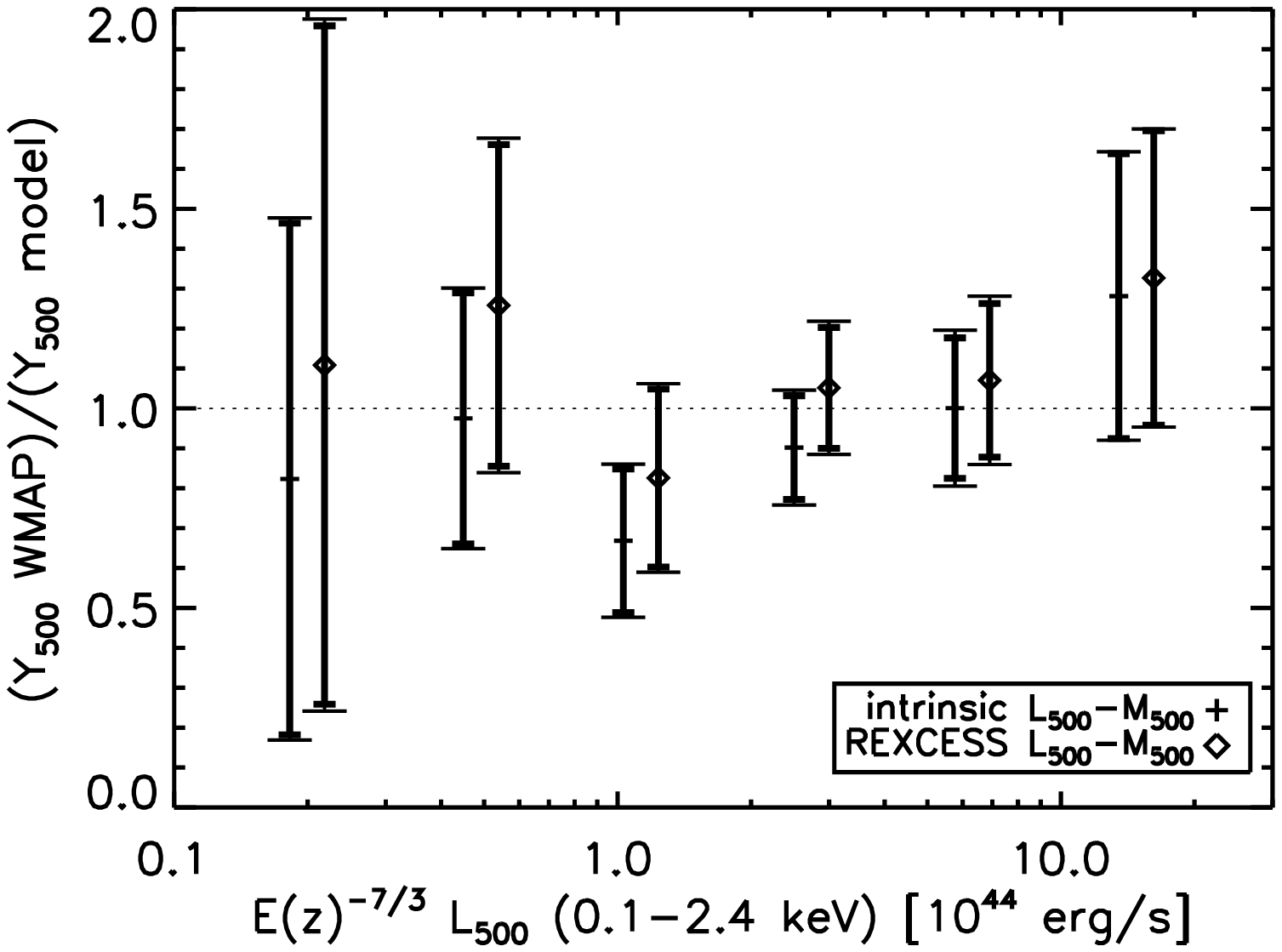}
   \caption{{\it Left:} Bin averaged SZ flux from a sphere of radius $r_{500}$ ($Y_{500}$) as a function of X-ray luminosity in a aperture of $r_{500}$ ($L_{500}$). The WMAP data (red diamonds and crosses), the SZ cluster signal expected from the X--ray based model  (blue stars) and the combination of the \YM\ and \LM\ relations (dash and dotted dashed lines) are given for two analyses, using respectively the intrinsic \LM\ and the \rexcess\ \LM\ relations.  As expected, the data points do not change significantly from one case to the other showing that the $Y_{500}$--$L_{500}$ relation is rather insensitive to the finer details of the underlying \LM\  relation.
    {\it Right:} Ratio of data points to model for the two analysis. The points for the analysis undertaken with the intrinsic \LM\  are shifted to lower luminosities by 20\% for clarity.}
    \label{fig:cy_lx_relation}
    \end{figure*}

    \begin{table*}
     \caption[]{Fitted parameters for the observed $Y_{SZ}$-$L_{500}$ relation. The X-ray based model gives $Y^{*L}_{500}=0.89|1.07 \times 10^{-3} \, \left ({h/0.719} \right )^{-5/2} {\rm arcmin}^2$, \\ $\alpha^L_Y=1.11|1.04$ and $\beta^L_Y=2/3$ for the \rexcess\ and intrinsic \LM\ relation, respectively.}
        \label{fit_param_un_l}
       \begin{center}
       \begin{tabular}{cccc}
        \hline
        \hline
        $L_{500}-M_{500}$  &$Y^{*L}_{500}$ [$10^{-3} \, \left ({h/0.719} \right )^{-2} {\rm arcmin}^2 $] & $\alpha^L_Y$ & $\beta^L_Y$ \\
         \hline
           \rexcess\  &   $ 0.92 \pm 0.08 \, {\rm stat} \; [\pm 0.10 \, {\rm tot} ]  $ & 1.11 (fixed) & 2/3 (fixed) \\
           &  $ 0.88 \pm 0.10 \, {\rm stat} \; [ \pm 0.12 \, {\rm tot} ] $ & $1.19 \pm 0.10 \, {\rm stat} \; [ \pm 0.10 \, {\rm tot}]$ & 2/3 (fixed) \\
            & $ 0.90 \pm 0.13 \, {\rm stat} \; [ \pm 0.16 \, {\rm tot} ] $ & 1.11 (fixed) &  $1.05 \pm 2.18 \, {\rm stat} \; [ \pm 2.25 \, {\rm tot} ]$\\
          \hline
           Intrinsic & $ 0.95 \pm 0.09 \, {\rm stat} \; [ \pm 0.11 \, {\rm tot} ] $ & 1.04 (fixed) & 2/3 (fixed) \\
          &    $ 0.89 \pm 0.10 \, {\rm stat} \; [ \pm 0.12 \, {\rm tot} ] $ & $1.19 \pm 0.10 \, {\rm stat} \; [ \pm 0.10 \, {\rm tot}]$ & 2/3 (fixed) \\
           &   $ 0.89 \pm 0.13 \, {\rm stat} \; [ \pm 0.16 \, {\rm tot} ] $ & 1.04 (fixed) &  $2.06 \pm 2.14 \, {\rm stat} \; [ \pm 2.21 \, {\rm tot}]$\\
         \hline
      \end{tabular}
       \end{center}
  \end{table*}

 The choice of the \LM\  relation  has an effect both on the estimated $\LX$, $\Mv$ and $\Yv$ values and on the expectation for  the SZ signal from the NORAS/REFLEX clusters. However, for a cluster of given luminosity measured  a given aperture, $\LX$ depends  weakly on the exact value of $r_{500}$ due to the steep drop of X--ray emission with  radius. As a result, and although $\LX$ and $\Mv$ (or equivalently $\Rv$) are determined jointly in the iterative procedure described in Sec.~\ref{rass}, changing the underlying  \LM\  relation mostly impacts the $\Mv$ estimate:  $\LX$ is essentially unchanged (median difference of  $\sim 0.8\%$) and the difference in $\Mv$ simply reflects the difference between the relations at fixed  luminosity. This has an impact on the measured  $Y_{500}$ via  the value of $r_{500}$ (the profile shape being fixed) but  the effect is also  small  ($<1\%$). This is due to the rapidly converging nature of the $Y_{SZ}$ flux (see Fig.~11 of \citealt{arnaud09}). On the other hand, all results that depend directly on $\Mv$, namely   the derived \YM\ relation or the model  value for each cluster, that  varies as $\Mv^{5/3}$ (Eq.~\ref{ym_ss_rel}), depend sensitively on the \LM\  relation.  $\Mv$ derived from the intrinsic relation is higher, an effect increasing with decreasing  cluster luminosity (see Fig. B2 of  \citealt{2009A&A...498..361P}).
 
\subsection{Other possible sources of uncertainty}

The analysis presented in this paper has been performed on the entire NORAS/REFLEX cluster sample without removal of clusters hosting radio point sources. To investigate the impact of the point sources on our result, we have cross-correlated the NVSS~\citep{condon98} and SUMMS~\citep{mauch03} catalogues with our cluster catalogue. We conservatively removed from the analysis all the clusters hosting a total radio flux greater than 1~Jy within $5 \times r_{500}$. This leaves 328 clusters in the catalogue, removing the measurements with large uncertainties visible in Figure~\ref{fig:cy_lx_raw_binned} left. We then performed the full analysis on these 328 objects up to the fitting of the scaling laws, finding that the impact on the fitted values is marginal. For example, for the \rexcess\ case, the normalisation of the \YM\ relation decreases from 1.60 to 1.37 (1.6 statistical $\sigma$) and the slope changes from 1.79 to 1.64 (1 statistical $\sigma$). The statistical errors on these parameters decrease respectively from 0.14 to 0.30 and from 0.15 to 0.40 due to the smaller number of remaining clusters in the sample.

The detection method does not take into account superposition effects along the line of sight, a drawback that is inherent to any SZ observation. Thus we cannot fully rule out that our flux estimates are not partially contaminated by low mass systems surrounding the clusters of our sample. Numerical simulations give a possible estimate of the contamination:  \cite{hallman2007} suggest that low-mass systems and unbound gas may contribute up to $16.3\%^{+7\%}_{-6.4\%}$ of the SZ signal. This would lower our estimated cluster fluxes by $\sim 1.5 \sigma$.

\section{The $Y_{\rm SZ}$-$L_{500}$ relation}
\label{yl_relation}

\subsection{WMAP SZ measurements vs. X--ray model}
\label{Szmes_Xray}

We first consider bin averaged data, focusing on the luminosity range $\LX \gtrsim10^{43}$\,ergs/s where the SZ signal is significantly detected (Fig.~\ref{fig:cy_lx_raw_binned} right). The left panel of  Fig.~\ref{fig:cy_lx_relation} shows $Y_{500}$ from a sphere of radius $\Rv$ as a function of $L_{500}$, averaging the data in six equally--spaced logarithmic bins in X--ray luminosity.  Both quantities are  scaled according to their expected redshift dependence.    The results are presented for the analyses based on the \rexcess\ (red diamonds) and intrinsic (red crosses) \LM\ relations. For the reasons discussed in Sec.~\ref{mal_bias}, the derived  data points do not differ significantly between the two analyses  (Fig.~\ref{fig:cy_lx_relation} left), confirming that the measured \YL\  relation is insensitive to the finer details of the underlying \LM\ relation. 
\begin{figure}
\centering
\includegraphics[scale=0.5]{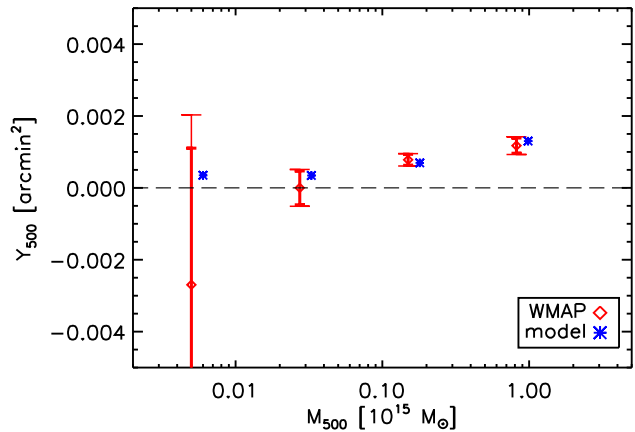}
 \caption{Estimated SZ flux $Y_{500}$ (in a sphere of radius $r_{500}$) as a function of the mass $M_{500}$ averaged in 4 mass bins. Red diamonds are the WMAP data. Blue stars correspond to the X-ray based model predictions and are shifted to higher masses by 20\% for clarity. The model is in very good agreement with the data.}
\label{fig:cy_m_relation}
\end{figure}

   \begin{figure*}[t]
   \centering
   \includegraphics[scale=0.5]{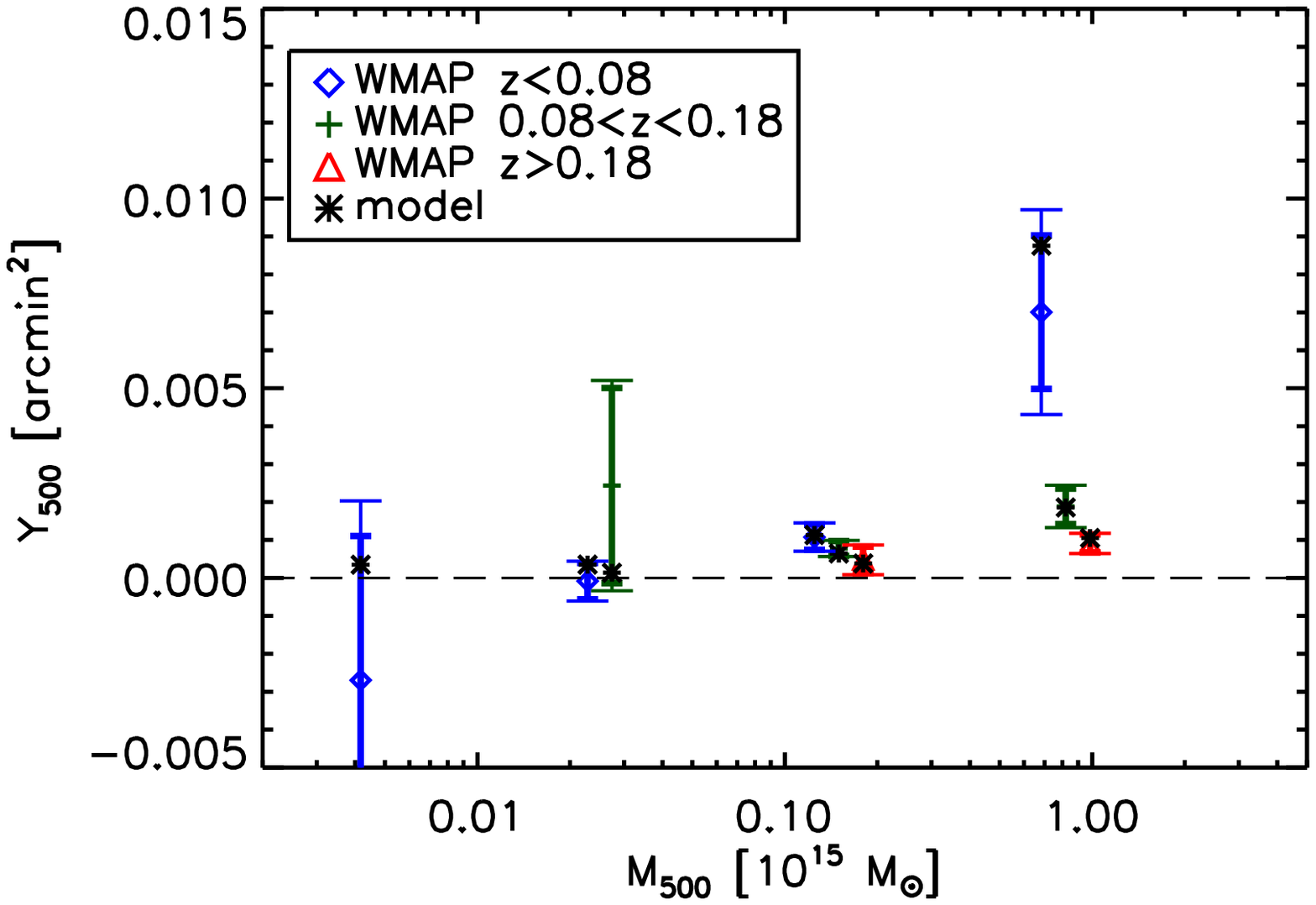}
    \includegraphics[scale=0.5]{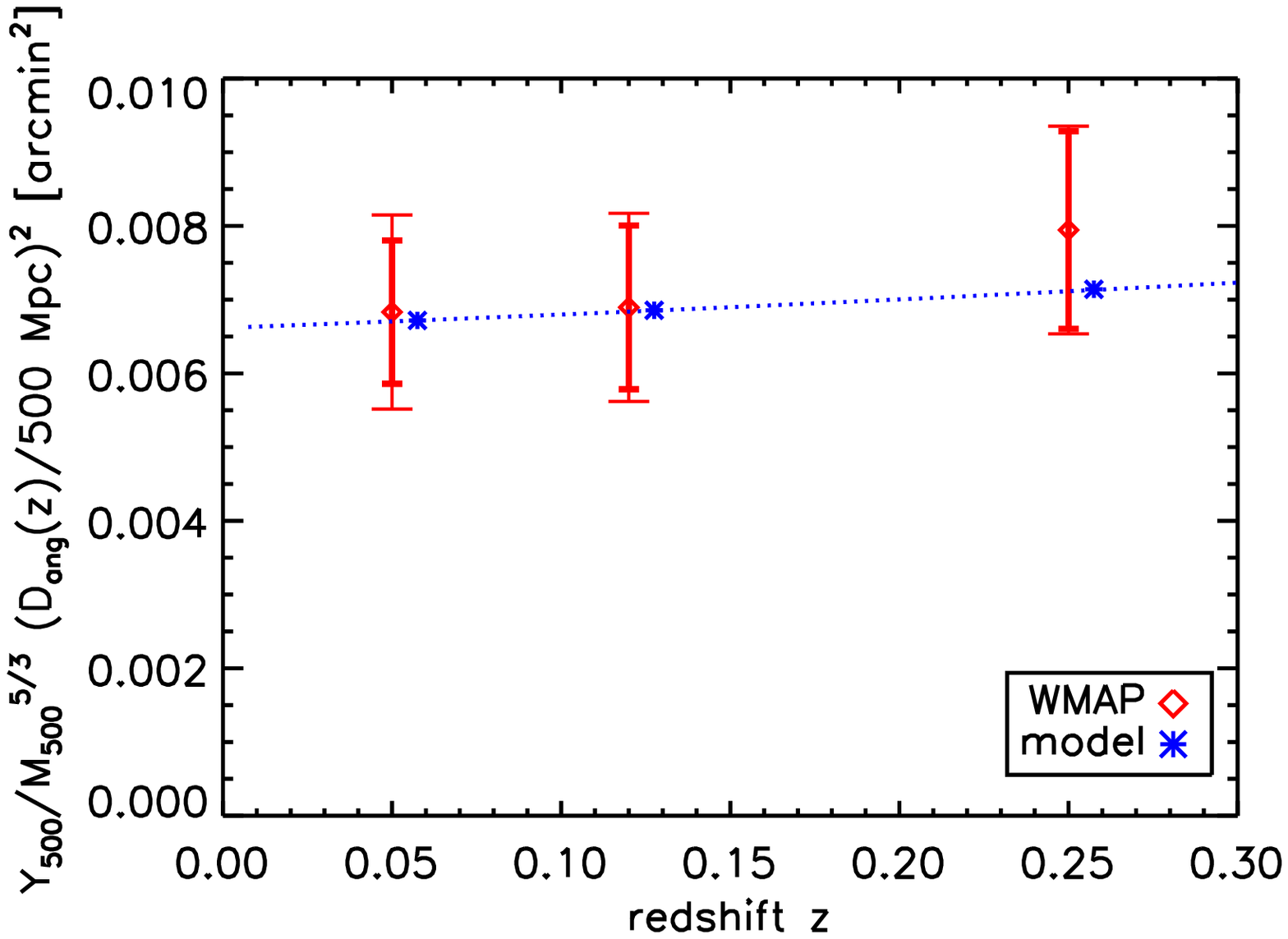}
        \caption{Evolution of the $Y_{500}$-$M_{500}$ relation. {\it Left:} The WMAP data from Fig.~\ref{fig:cy_m_relation}  are divided into three redshift bins: z$<$0.08 (blue diamonds), 0.08$<$z$<$0.18 (green crosses), z$>$0.18 (red triangles). We observe the expected trend: at fixed mass, $Y_{500}$ decreases with redshift. This redshift dependence is mainly due to the angular distance ($Y_{500} \propto D_{ang}(z)^{-2}$). The stars give the prediction of the model. {\it Right:} We divide $Y_{500}$ by $M_{500}^{5/3} D_{ang}(z)^{-2}$ and plot it as a function of $z$ to search for evidence of evolution in the $Y_{500}$-$M_{500}$ relation. The thick bars give the 1 $\sigma$ statistical errors from WMAP data. The thin bars give the total 1 sigma errors.}
    \label{fig:evolution}
    \end{figure*}

We also apply the same averaging procedure to the model $Y_{500}$ values derived for each cluster  in Sec.~\ref{cluster_model}. The expected values for the same luminosity bins are plotted as stars in the left-hand hand panel of  Fig.~\ref{fig:cy_lx_relation}.  The \YL\ relation expected from the combination of the \YM\ (Eq.~\ref{ym_ss_rel}) and \LM\ (Eq.~\ref{lx_m500}) relations is superimposed to guide the eye. 
The right-hand panel of Fig.~\ref{fig:cy_lx_relation} shows the ratio between the measured data points and those expected from the model.  As discussed in Sec.~\ref{mal_bias},  the  model values depend on the assumed \LM\  relation. The difference is maximum  in the lowest luminosity bin where the intrinsic relation yields $\sim 40\%$ higher value than the \rexcess\ relation  (Fig.~\ref{fig:cy_lx_relation} left panel). The  SZ model prediction and the data are in good agreement, but  the agreement is better  when the \rexcess\  \LM\  is used in the analysis (Fig.~\ref{fig:cy_lx_relation} right panel). This is expected if indeed the  agreement is real and the effective Malmquist bias for the NORAS/REFLEX sample  is not negligible and is similar to  that of the \rexcess.

\begin{figure*}[t]
\centering
\includegraphics[scale=0.5]{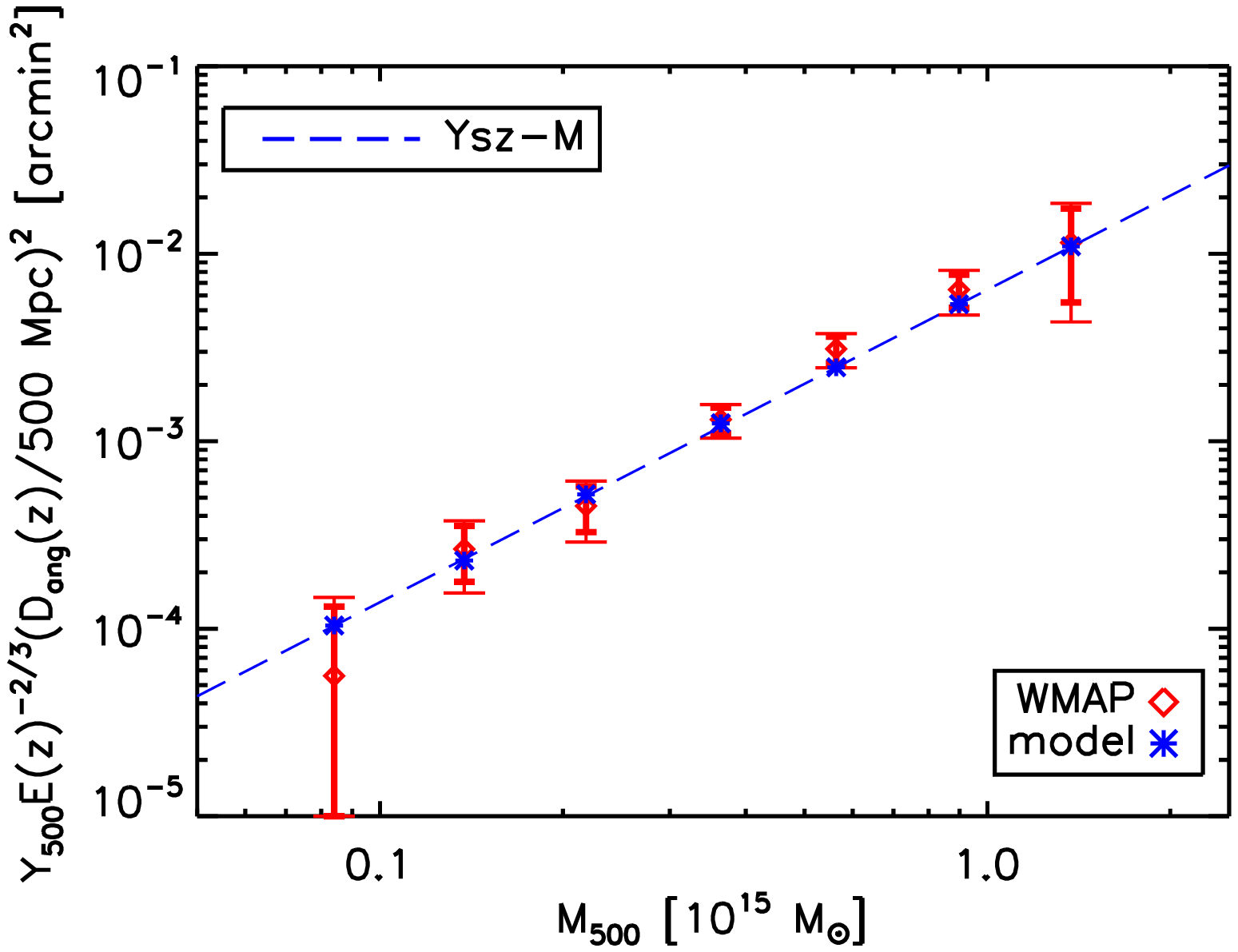}
\includegraphics[scale=0.5]{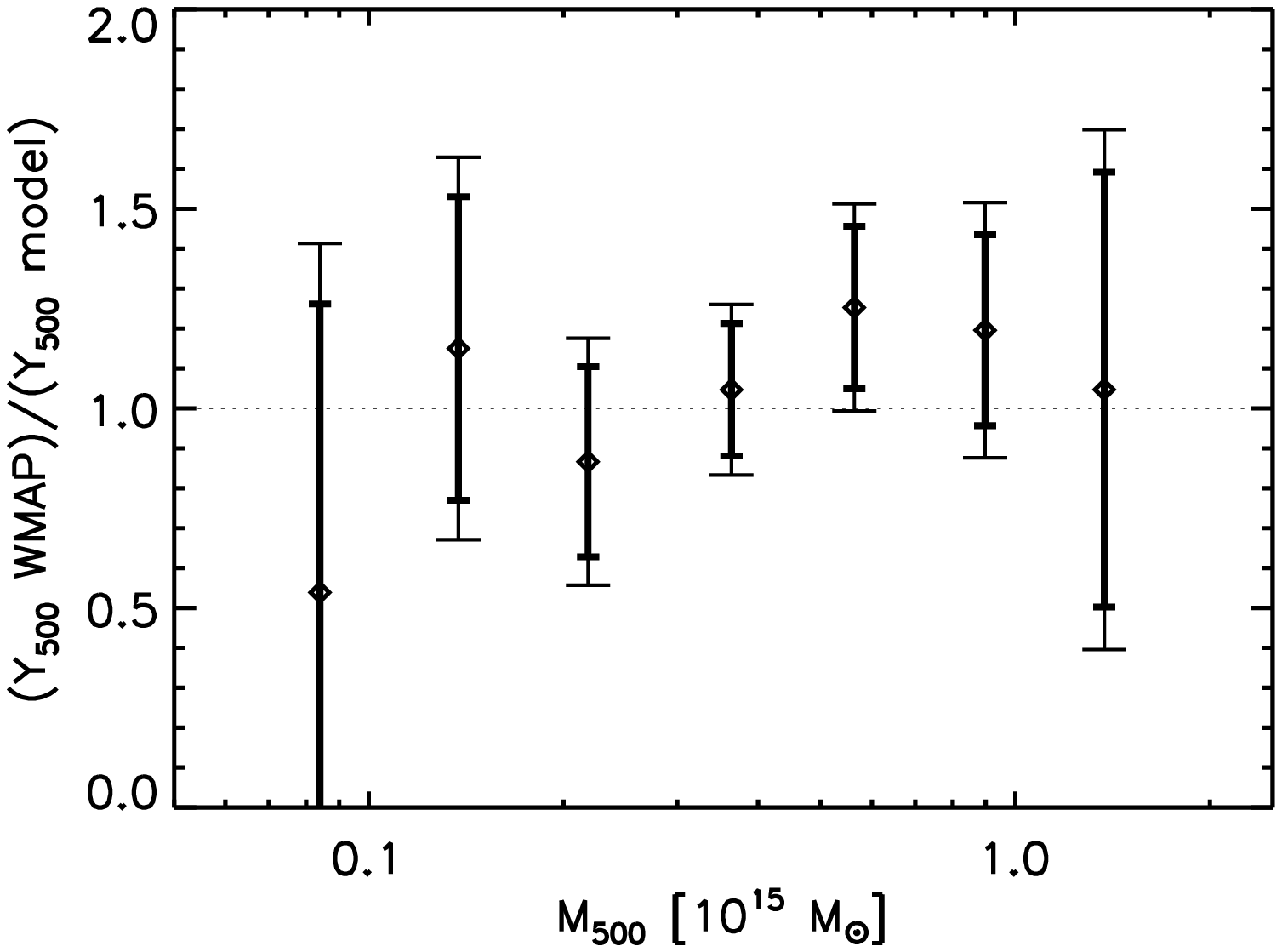}
 \caption{{\it Left:} Zoom on the $> 5 \times 10^{13} M_\odot$ mass range  of the $Y_{500}-M_{500}$ relation shown in Fig.~\ref{fig:cy_m_relation}. The data points and model stars are now scaled with the expected redshift dependence and are placed at the mean mass of the clusters in each bin. {\it Right:} Ratio between data and model.}
\label{fig:cy_m_loglog}
\end{figure*}

\subsection{$Y_{500}$-$L_{500}$ relation fit}
\label{yl_fit_sec}

Working now with the individual flux measurements, $Y_{500}$, and $L_{500}$ values,  we fit  an  \YL\ relation of the form:
     \begin{equation}
      Y_{500} = Y^{*L}_{500} \; \left ({E(z)^{-7/3} L_{500} \over 10^{44} h^{-2} {\rm erg/s}} \right )^{\alpha^L_Y} \; E(z)^{\beta^L_Y} \; \left ({D_{ang}(z) \over 500 \, {\rm Mpc}} \right )^{-2}
     \end{equation}
using the statistical error on $Y_{500}$ given by the  multifrequency matched filter.   The total  error is estimated by Monte Carlo (see Sec.~\ref{disp_err}) but is dominated by the statistical error.   The results are presented in Table~\ref{fit_param_un_l}. As already stated in Sec.~\ref{Szmes_Xray}, the fitted values depend only weakly on the choice of  \LM\ relation.

\section{The $Y_{\rm SZ}$--$M_{500}$ relation and its evolution}
\label{ym_relation}

 In this Section, we study the  mass and redshift dependence of the SZ signal  and check it against the X--ray based  model. Furthermore, we fit the \YM\ relation and compare it with the X--ray predictions. 

\subsection{Mass dependence and redshift evolution}

Figure~\ref{fig:cy_m_relation} shows the bin averaged SZ flux measurement as a function of mass compared to the X--ray based model prediction. As expected, the SZ cluster flux increases as a function of mass and is compatible with the model. In order to study the behaviour of the SZ flux with redshift, we subdivide each of the four mass bins into three redshift bins corresponding to the following ranges: $z<0.08$, $0.08<z<0.18$, $z>0.18$. The result is shown in the left panel of Fig.~\ref{fig:evolution}.
In a given mass bin the SZ flux decreases with redshift, tracing the $D_{ang}(z)^{-2}$ dependence of the flux. In particular, in the highest mass bin ($10^{15} M_\odot$), the SZ flux decreases from 0.007 to $0.001 \, {\rm arcmin^2}$ while the redshift varies from $z<0.08$ to $z>0.18$. The mass and the redshift dependence are in good agreement with the model (stars) described in Sec.~\ref{cluster_model}.

Since the $D_{ang}(z)^{-2}$ dependence is the dominant effect in the redshift evolution, we multiply $Y_{500}$ by $D_{ang}(z)^{2}$ and divide it by the self-similar mass dependence $M_{500}^{5/3}$. The expected self-similar behaviour of the new quantity $Y_{500}D_{ang}(z)^{2}/M_{500}^{5/3}$ as a function of redshift is $E(z)^{2/3}$ (see Eq.~\ref{ym_ss_rel}). The right panel of Fig.~\ref{fig:evolution} shows $Y_{500}D_{ang}(z)^{2}/M_{500}^{5/3}$ as a function of redshift for the three redshift bins $z<0.08$, $0.08<z<0.18$, $z>0.18$. The points have been centered at the average value of the cluster redshifts in each bin. The model is displayed as blue stars. Since the model has a self-similar redshift dependence and $E(z)^{2/3}$ increases only by a factor of  5\% over the studied redshift range, the model stays nearly constant. The blue dotted line is plotted through the model and varies as $E(z)^{2/3}$. The data points are in good agreement with the model, but clearly, the redshift leverage of the sample is insufficient  to put strong constraints on the evolution of the scaling laws.
   \begin{table*}[t]
     \caption[]{Fitted parameters for the observed $Y_{SZ}$-$M_{500}$ relation. The X-ray based model gives $Y^*_{500}=1.54 \times 10^{-3} \, \left ({h/0.719} \right )^{-5/2} {\rm arcmin}^2$, $\alpha_Y=5/3$ and $\beta_Y=2/3$.}
        \label{fit_param_un}
       \begin{center}
       \begin{tabular}{cccc}
        \hline
        \hline
       \LM\ relation &  $Y^*_{500}$ [$10^{-3} \, \left ({h/0.719} \right )^{-2} \, {\rm arcmin}^2$] & $\alpha_Y$ & $\beta_Y$ \\
         \hline
        \rexcess\  &     $1.60 \pm 0.14 \, {\rm stat} \; [ \pm 0.19 \, {\rm tot} ] $ & 5/3 (fixed) & 2/3 (fixed) \\
           &  $ 1.60 \pm 0.15 \, {\rm stat} \; [ \pm 0.19 \, {\rm tot} ] $ & $1.79 \pm 0.15 \, {\rm stat} \; [ \pm 0.17 \, {\rm tot}]$ & 2/3 (fixed) \\
         &    $ 1.57 \pm 0.23 \, {\rm stat} \; [ \pm 0.29 \, {\rm tot} ] $ & 5/3 (fixed) &  $1.05 \pm 2.18 \, {\rm stat} \; [ \pm 2.52 \, {\rm tot}]$\\
        \hline
      intrinsic       &    $ 1.37 \pm 0.12 \, {\rm stat} \; [ \pm 0.17 \, {\rm tot} ] $ & 5/3 (fixed) & 2/3 (fixed) \\
         &    $ 1.36 \pm 0.13 \, {\rm stat} \; [ \pm 0.17 \, {\rm tot} ] $ & $1.91 \pm 0.16 \, {\rm stat} \; [ \pm 0.18 \, {\rm tot}]$ & 2/3 (fixed) \\
              &  $ 1.28 \pm 0.19 \, {\rm stat} \; [ \pm 0.24 \, {\rm tot} ] $ & 5/3 (fixed) &  $2.06 \pm 2.14 \, {\rm stat} \; [ \pm 2.48 \, {\rm tot}]$\\
        \hline
     \end{tabular}
       \end{center}
  \end{table*}

We now focus on the mass dependence of the relation. We scale the SZ flux with the expected redshift dependence and plot it as a function of mass. The result is shown in Fig.~\ref{fig:cy_m_loglog} for the high mass end. The figure shows a very good agreement between the data points and the model, which is confirmed by fitting  the relation to the individual SZ flux measurements (see next Section).

\subsection{$Y_{500}$--$M_{500}$ relation fit}
\label{ym_fit_sec}

Using the individual $Y_{500}$ measurements  and $M_{500}$  estimated from the X--ray luminosity,  we fit a relation of the form:
     \begin{equation}
      Y_{500} = Y^*_{500} \; \left ({M_{500} \over 3 \times 10^{14} h^{-1} M_\odot} \right )^{\alpha_Y} \; E(z)^{\beta_Y} \; \left ({D_{ang}(z) \over 500 \, {\rm Mpc}} \right )^{-2}
     \end{equation}
     The results are presented in Table~\ref{fit_param_un}  for the analysis undertaken using  the  \rexcess\ and that using the  intrinsic  \LM\ relation. The pivot mass $3 \times 10^{14} h^{-1} M_\odot$, close to that  used by \citet{arnaud09}, is slightly larger than the average mass of the sample ($2.8|2.5 \, \times 10^{14} M_\odot$ for the \rexcess|intrinsic \LM\ relation, respectively). We use a non-linear least-squares fit built on a gradient-expansion algorithm (IDL curvefit function). In the fitting procedure, only the statistical errors given by the matched multifilter ($\sigma_{Y_{500}}$) are taken into account. The total  errors on the final fitted parameters, taking into account uncertainties in  X--ray properties,  are estimated by Monte Carlo as described in Sec.~\ref{err_bud}. 

We first discuss the results obtained using the \rexcess\ \LM\ relation, which is expected to be close  to the optimal case (see discussion in Sec.~\ref{mal_bias}).
First, we keep the mass and redshift dependence fixed to the self-similar expectation ($\alpha_{\rm Y}=5/3$, $\beta_{\rm Y}=2/3$) and we fit only the normalisation. We obtain $Y^*_{500} = 1.60 \times 10^{-3} \, \left ({h/0.719} \right )^{-2} \, {\rm arcmin}^2$, in agreement with the X-ray prediction $Y^*_{500}=1.54 \times 10^{-3} \, \left ({h/0.719} \right )^{-5/2}  \, {\rm arcmin}^2$ (at $ 0.4 \sigma$). Then, we relax the constraint on $\alpha_{\rm Y}$ and fit the normalisation and mass dependence at the same time. We obtain a value for $\alpha_{\rm Y}=1.79$, slightly greater than the self-similar expectation (5/3) by $0.8\,\sigma$. To study the redshift dependence of the effect, we fix the mass dependence to $\alpha_{\rm Y}=5/3$ and fit $Y^*_{500}$ and $\beta_{\rm Y}$ at the same time. We obtain a somewhat stronger evolution $\beta_{\rm Y}=1.05$ than the self-similar expectation (2/3). The difference, however, is not significant ($0.2 \sigma$). As already mentioned above (see also Fig.~\ref{fig:evolution} right), the redshift leverage is too small to get interesting constraints on $\beta_{\rm Y}$.

As cluster mass estimates depend on the assumption of the underlying \LM\ relation, so does the derived \YM\ relation as well. However, the effect is small. The normalisation  is shifted from $ \left (1.60 \pm 0.14 \, {\rm stat} \; [ \pm 0.19 \, {\rm tot} ] \right ) \, 10^{-3} \, {\rm arcmin}^2$ to $ \left (1.37 \pm 0.12 \, {\rm stat} \; [ \pm 0.17 \, {\rm tot} ] \right ) \, 10^{-3} \, {\rm arcmin}^2$ when using the intrinsic \LM\ relation. The difference is less than two statistical sigmas, and for  the mass exponent, it is less than one.

\section{Discussion and conclusions}
\label{discussion}

In this paper we have investigated the SZ effect and its scaling with mass and X-ray luminosity using WMAP 5-year data of the largest published X-ray-selected cluster catalogue to date, derived from the combined NORAS and REFLEX samples. Cluster SZ flux estimates were made using an optimised multifrequency matched filter. Filter optimisation was achieved through priors on the pressure distribution (i.e., cluster shape) and the integration aperture (i.e., cluster size). The pressure distribution is assumed to follow the universal pressure profile of \citet{arnaud09}, derived from X-ray observations of the representative local \rexcess\ sample. This profile is the most realistic available for the general population at this time, and has been shown to be in good agreement with recent high-quality SZ observations from SPT \citep{pla09}. Furthermore, our analysis takes into account the dispersion in the pressure distribution. The integration aperture is estimated from the $L_{500}-M_{500}$ relation of the same \rexcess\ sample. We emphasise that these two priors determine only the input spatial distribution of the SZ flux for use by the multifrequency matched filters; the priors give no information on the amplitude of the measurement. As the analysis uses minimal X-ray data input, the measured and X-ray predicted SZ fluxes are essentially independent.

We studied the $Y_{SZ}-L_X$ relation using both bin averaged analyses and individual flux measurements. The fits using individual flux measurements give quantitative results for calibrating the scaling laws.  The bin averaged analyses allow a direct quantitative check of SZ flux measurements versus X-ray model predictions based on the universal pressure profile derived by \citet{arnaud09} from \rexcess. An excellent agreement is found.

Using WMAP 3-year data, both \citet{2006ApJ...648..176L} and \citet{bie07} found that the SZ signal strength is lower than predicted given expectations from the X-ray properties of their clusters, concluding that that there is some missing hot gas in the intra-cluster medium. The excellent agreement between the SZ and X-ray properties of the clusters in our sample shows that there is in fact no deficit in SZ signal strength relative to expectations from X-ray observations. Due to the large size and homogeneous nature of our sample, and the internal consistency of our baseline cluster model, we believe our results to be robust in this respect.  We note that there is some confusion in the literature regarding the phrase `missing baryons'. The `missing baryons' mentioned by \citet{2007MNRAS.378..293A} in the WMAP 3-year data are missing with respect to the universal baryon fraction, but not with respect to the expectations from X-ray measurements. \citet{2007MNRAS.378..293A} actually found good agreement between the strength of the SZ signal and the X-ray properties of their cluster sample, a conclusion that agrees with our results. This good convergence between SZ direct measurements and X--ray data is an encouraging step forward for the prediction and interpretation of SZ surveys.

Using $L_{500}$ as a mass proxy, we also calibrated the \YM\ relation, finding a normalisation in excellent agreement with X-ray predictions based on the universal pressure profile, and a slope consistent with self-similar expectations. However, there is some indication that the slope may be steeper, as also indicated from the \rexcess\  analysis when using the best fitting empirical $\Mv$--$Y_{\rm X}$ relation \citep{arnaud09}. $M_{500}$ depends on the assumed $L_{500}-M_{500}$ relation, making the derived \YM\ relation sensitive to  Malmquist bias which we cannot fully account for in our analysis. However, we have shown that the effect of Malmquist bias on the present results is less than $2 \sigma$ (statistical).

Regarding evolution, we have shown observationally that the SZ flux is indeed sensitive to the angular size of the cluster through the diameter distance effect. For a given mass, a low redshift cluster has a bigger integrated SZ flux than a similar system at high redshift, and the redshift dependence of the integrated SZ flux is dominated by the angular diameter distance ($\propto D_{ang}^2(z)$ ). However, the redshift leverage of the present cluster sample is too small to put strong contraints on the evolution of the \YL\ and \YM\  relations. We have nevertheless checked that the observed evolution is indeed compatible with the self-similar prediction.

In this analysis, we have compensated for the poor sensitivity and resolution of the WMAP experiment (regarding SZ science) with the large number of known ROSAT clusters, leading to self-consistent and robust results. We expect further progress using upcoming Planck all-sky data. While Planck will offer the possibility of detecting the clusters used in this analysis to higher precision, thus significantly reducing the uncertainty on individual measurements, the question of evolution will not be answered with the present  RASS sample due to its limited redshift range. A complementary approach will thus be to obtain new high sensitivity SZ observations of a smaller sample. The sample must be representative, cover a wide mass range, and extend to higher z (e.g., XMM-Newton follow-up of samples drawn from Planck and ground based SZ surveys). This would deliver efficient constraints not only on the normalisation and slope  of the $Y_{SZ}-L_X$ and $Y_{SZ}-M$ relations, but also their evolution, opening the way for the use of SZ surveys for precision cosmology.

\begin{acknowledgements}
The authors wish to thank the anonymous referee for useful comments. J.-B. Melin thanks R. Battye for suggesting introduction of the $h$ dependance into the presentation of the results. The authors also acknowledge the use of the HEALPix package~(\cite{2005ApJ...622..759G}) and of the Legacy Archive for Microwave Background Data Analysis (LAMBDA). Support for LAMBDA is provided by the NASA Office of Space Science.  We also acknowledge use of the Planck Sky Model, developed by the Component Separation Working Group (WG2) of the Planck Collaboration, for the estimation of the radio source flux in the clusters and for the development of the matched multifilter, although the model was not directly used in the present work.
\\
\end{acknowledgements}

\appendix

\section{SZ flux definitions}
\label{sz_ref}

\begin{table}[b]
   \caption[]{Equivalence of SZ flux definitions}
   \label{flux_def}
        \begin{center}
        \begin{tabular}{|c|c|c|c|c|c|}
         \hline
          n & 1 & 2 & 3 & 5 & 10 \\
          \hline
          \hline
          $Y_{nr500}/Y_{500}$ & 1 & 1.505 & 1.690 & 1.814 & 1.873 \\
          \hline
          $Y_{nr500}/Y^{cyl}_{nr500}$ & 0.827 & 0.930 & 0.963 & 0.986 & 0.997 \\
          \hline
        \end{tabular}
        \end{center}
   \end{table}

In this Appendix, we give the definitions of SZ fluxes we used. Table~\ref{flux_def} gives the equivalence between them. In this paper, we mainly use $Y_{500}$ as the definition of the SZ flux. This flux is the integrated  SZ flux from a sphere of radius $r_{500}$.
It can be related to $Y_{nr500}$, the flux from a sphere of radius $n \times r_{500}$ by integrating over the cluster profile:

     \begin{equation}
      Y_{nr500} = Y_{500} {\int_0^{nr_{500}} dr P(r) 4 \pi r^2 \over \int_0^{r_{500}} dr P(r) 4 \pi r^2}
     \end{equation}

\noindent where $P(r)$ is given by Eq.~\ref{cluster_profile}. The ratio $Y_{nr500}/Y_{500}$ is given in Table~\ref{flux_def} for $n=1,\,2\,,3\,,5\,,10$.

In practice, an experiment does not directly measure $Y_{500}$ but the SZ signal of a cluster integrated along the line of sight and within an angular aperture. This corresponds to the Compton parameter integrated over a cylindrical  volume. In Sec.~\ref{sz_flux}, we estimate $Y^{cyl}_{5r500}$, the flux from  a cylinder of aperture radius $5 \times r_{500}$ using the matched multifilter. Given the cluster profile, we can derive $Y_{nr500}$ from $Y^{cyl}_{nr500}$:

\begin{equation}
   Y^{cyl}_{nr500}=Y_{nr500}{ \int_0^\infty dr \int_{r \sin \theta < nr_{500}} d \theta \, P(r) 2 \pi r^2 \over \int_0^{nr_{500}} dr P(r) 4 \pi r^2}
\end{equation}

\noindent The ratio $Y_{nr500}/Y^{cyl}_{nr500}$ is given in Table~\ref{flux_def} for  $n=1,\,2\,,3\,,5\,,10$. In the paper, we calculate $Y_{500}$ from $Y_{500}=0.986/1.814 \times Y^{cyl}_{5r500}$.

\section{SZ units conversion} 
\label{sz_conv}

In this Appendix, we provide the numerical factor needed for the SZ flux units conversion and derive the relation between the recently introduced $Y_X$ parameter and the SZ flux $Y_{SZ}$. The latter will allow readers to easily convert between SZ fluxes given in this paper and those reported in other publications.

Given the definition of SZ flux: 
\begin{equation} 
   Y^{cyl}_{nr500}=\int_{\Omega_{nr500}}  d\Omega \, y
\end{equation} 
\noindent where $\Omega_{nr500}$ is the solid angle covered by $n \times r_{500}$, and the fact that the Compton parameter $y$ is unitless, the observational units for the SZ flux are those of a solid angle and usually given in ${\rm arcmin}^2$.

The SZ flux can be also computed in units of ${\rm Mpc}^2$ and the conversion is given by 
\begin{eqnarray} 
   Y_{SZ} [{\rm Mpc}^2] & = &  60^{-2} \, \left ( {\pi \over 180} \right )^2 \, Y_{SZ} [{\rm arcmin}^2]  \, \left ( D_{ang}(z) \over 1 \, {\rm Mpc} \right )^2  \nonumber\\
   & = &  8.46 \times 10^{-8} \, Y_{SZ} [{\rm arcmin}^2]  \, \left ( D_{ang}(z) \over 1 \, {\rm Mpc} \right )^2
\end{eqnarray} 
\noindent where $D_{ang}(z)$ is the angular distance to the cluster. \\

The X-ray analogue of the integrated SZ  Comptonisation parameter is $Y_X=M_{gas,500}T_X$ whose natural units are $ {\rm M}_\odot \, {\rm keV}$, where $M_{gas,500}$ is the gas mass in $r_{500}$ and $T_X$ is the {\it spectroscopic} temperature excluding the central $0.15\,r_{500}$ region \citep{kra06}. To convert between $Y_{SZ}$ and $Y_X$, we first have
\begin{equation} 
   Y_{SZ} [{\rm Mpc}^2] = {\int_0^{r_{500}} dr \, \sigma_T \, {T_e(r) \over m_e c^2} \, n_e(r) 4 \pi r^2} 
\end{equation} 
\noindent where $\sigma_T$ is the Thomson cross section (in ${\rm Mpc}^2$), $m_e c^2$ the electron mass (in ${\rm keV}$), $T_e(r)$ the electronic temperature (in ${\rm keV}$) and $n_e(r)$ the electronic density.
By assuming that the gas temperature $T_g(r)$ is equal to the electronic temperature $T_e(r)$ and writing the gas density as $\rho_g(r)=\mu_e m_p n_e(r)$, where $m_p$ is the proton mass and $\mu_e=1.14$ the mean molecular weight per free electron, one obtains:
\begin{eqnarray} 
   Y_{SZ} [{\rm Mpc}^2] & = & {\sigma_T \over m_e c^2} {1 \over \mu_e m_p} {\int_0^{r_{500}} dr \, \rho_g(r) \, T_g(r) 4 \pi r^2} \nonumber\\
  & = & C_{{\rm XSZ}} \, M_{gas,500} \, T_{MW} = A \, C_{{\rm XSZ}} \, Y_X 
\end{eqnarray} 
where, as in~\cite{arnaud09}, we defined 
\begin{equation} 
  C_{{\rm XSZ}}={\sigma_T \over m_e c^2} {1 \over \mu_e m_p}=1.416 \times 10^{-19}  \; {{\rm Mpc}^2 \over {\rm M}_\odot \, {\rm keV}}
\end{equation} 
The {\it mass weighted} temperature is defined as: 
\begin{equation} 
T_{MW}= { \int_0^{r_{500}} dr \, \rho_g(r) \, T_g(r) 4 \pi r^2 \over \int_0^{r_{500}} dr \, \rho_g(r) 4 \pi r^2 } 
\end{equation} 
\noindent and the factor $A=T_{MW}/T_X$ takes into account for the difference between mass weighted and spectroscopic average temperatures. \cite{arnaud09} find $A \sim 0.924$.
\end{document}